\long\def\symbolfootnote[#1]#2{\begingroup%
\def\thefootnote{\fnsymbol{footnote}}\footnote[#1]{#2}\endgroup} 
\def\aj{AJ}
\def\araa{ARA\&A}
\def\apj{ApJ}
\def\apjl{ApJ}
\def\apjs{ApJS}
\def\mnras{MNRAS}

\newcommand{\dr}{{\rm d}}
\newcommand{\mg}{M_{\rm g}}
\newcommand{\ms}{M}
\newcommand{\msun}{{\rm M}_\odot}

\newcommand{\myr}{{\rm Myr}}

\newcommand{\ns}{N}

\newcommand{\nslow}{N_{\rm low}}
\newcommand{\nsup}{N_{\rm up}}
\newcommand{\pc}{{\rm pc}}

\newcommand{\rn}{R_0}

\newcommand{\phic}{\phi_{\rm cl}(N)}
\newcommand{\rh}{r_{\rm h}}
\newcommand{\rhn}{r_{\rm h,1}}
\newcommand{\sigs}{\sigma}
\newcommand{\rhos}{\rho}
\newcommand{\rs}{\rh}
\newcommand{\rg}{r_{\rm h,g}}
\newcommand{\sig}{\Sigma}

\newcommand{\sigc}{\Sigma(R)}
\newcommand{\siglow}{\Sigma_{0,\rm low}}
\newcommand{\tilsiglow}{\tilde{\Sigma}_{0,\rm low}}
\newcommand{\sigmin}{\Sigma_{0,\rm min}}
\newcommand{\sigup}{\Sigma_{0,\rm up}}
\newcommand{\sigcn}{\Sigma_{\rm 0}}
\newcommand{\sigco}{\Sigma_{\rm 1}}
\newcommand{\trh}{\tau_{\rm rh}}

\documentclass[useAMS,usenatbib, times]{mn2e}
\usepackage{amssymb,latexsym,graphicx,natbib,eufrak,times,amsmath}

\title[Do all stars  form in clusters?]
  {Do all stars in the solar neighbourhood form in clusters?\\
  A cautionary note on the use of the distribution of surface densities}
\author[M. Gieles, N. Moeckel \& C. Clarke]
  {Mark~Gieles, {Nickolas Moeckel} and Cathie J. Clarke\\
  Institute of Astronomy, University of Cambridge, Madingley Road, Cambridge, CB3 0HA, UK 
}
\date{Accepted 2012 July 9. Received 2012 July 09; in original form 2012 June 24}

\def\LaTeX{L\kern-.36em\raise.3ex\hbox{a}\kern-.15em
    T\kern-.1667em\lower.7ex\hbox{E}\kern-.125emX}

\begin{document}         

\maketitle
\begin{abstract}
Bressert et al. recently showed that the surface density distribution
of low-mass, young stellar objects (YSOs) in the solar neighbourhood is
approximately lognormal. The authors  conclude that the star
formation process is hierarchical and that only a small fraction of
stars form in dense star clusters. Here, we show that the peak and the
width of the density distribution is also what follows if all stars
form in bound clusters  which are not significantly affected by
the presence of gas and expand by two-body relaxation. The peak of
the surface density distribution is simply obtained from the typical
ages (few Myr) and cluster membership number (few hundred) typifying
nearby star forming regions. This result depends weakly on initial
cluster sizes, provided that they are sufficiently dense (initial half
mass radius of $\lesssim$ 0.3 pc) for dynamical evolution to be
important at an age of a few Myr. We conclude that the degeneracy of
the YSO surface density distribution complicates its use as a
diagnostic of the stellar formation environment.
\end{abstract}

\section{Introduction}
\label{sec:intro}
We do not know what fraction of stars form in dense star
clusters. Part of the problem is that there is no agreed definitions
of what `dense' means and what a star cluster
is \citep*{2010ARA&A..48..431P}.  In an attempt to shed light on this
situation \citet{2010MNRAS.409L..54B} studied a sample of young
stellar objects (YSOs) in the solar neighbourhood. They calculate the
surface density $\sig$ around each YSO by finding the distance to the
7$^{\rm th}$ nearest nearest neighbour, $d_7$, such that $\sig=6/(\pi
d_7^2)$. They find that the distribution of $\sig$ is roughly
lognormal with a peak at about $22\,\pc^{-2}$ and a dispersion of
0.85. Because YSOs are very young (of order 1 Myr) they conclude that
this distribution reflects the density distribution at the moment of
star formation.  They concluded that stars form in a broad and smooth
spectrum of surface densities and that only a small fraction of the
YSOs form in dense clusters.

In this paper we argue that the observed peak in the surface density
distribution of young stars (at around $20$ stars per square parsec)
is an expected outcome from a wide range of initial clustering
configurations.  In Section \ref{sec:peak} we argue that for `typical'
cluster scales in star forming regions 
\citep[$N \simeq 100$,][]{2003ARA&A..41...57L}
and young ages (about 1 Myr) of YSOs such a surface density
represents the outcome of dynamical evolution (i.e. two-body
relaxation) from a variety of plausible initial conditions.  In
Section \ref{sec:model} we flesh out this argument by considering the
factors that broaden this distribution (the range of surface densities
in a given cluster together with a realistic spectrum of cluster
membership number).  In Section \ref{sec:conclusions} we  present
a discussion and conclude that the observed surface density
distribution is exactly what one expects if the majority of stars are
born in clusters; the situation is however highly degenerate so that
it is not possible to use this distribution to place unique
constraints on the initial conditions.

\section{The peak of the cluster surface density distribution}
\label{sec:peak}
The evolution of self-gravitating clusters containing a realistic
stellar mass spectrum is well understood. The massive stars segregate
to the cluster core due to dynamical drag and initiate core collapse:
in the case of clusters numbering a few hundred stars, this occurs
after about 10 or 20 crossing
times \citep{1996MNRAS.279.1037G}. Subsequently the cluster expands
due to the transfer of energy from binaries and stellar mass-loss in
the core by two-body relaxation. This expansion approaches a
self-similar state in which the cluster expands more or less
homologously and with radius scaling as
$t^{2/3}$ \citep{1996MNRAS.279.1037G}. In this state the two-body
relaxation timescale tends to a fixed multiple of the cluster
age \citep{1965AnAp...28...62H}.

\begin{figure} 
\includegraphics[width=8.cm]{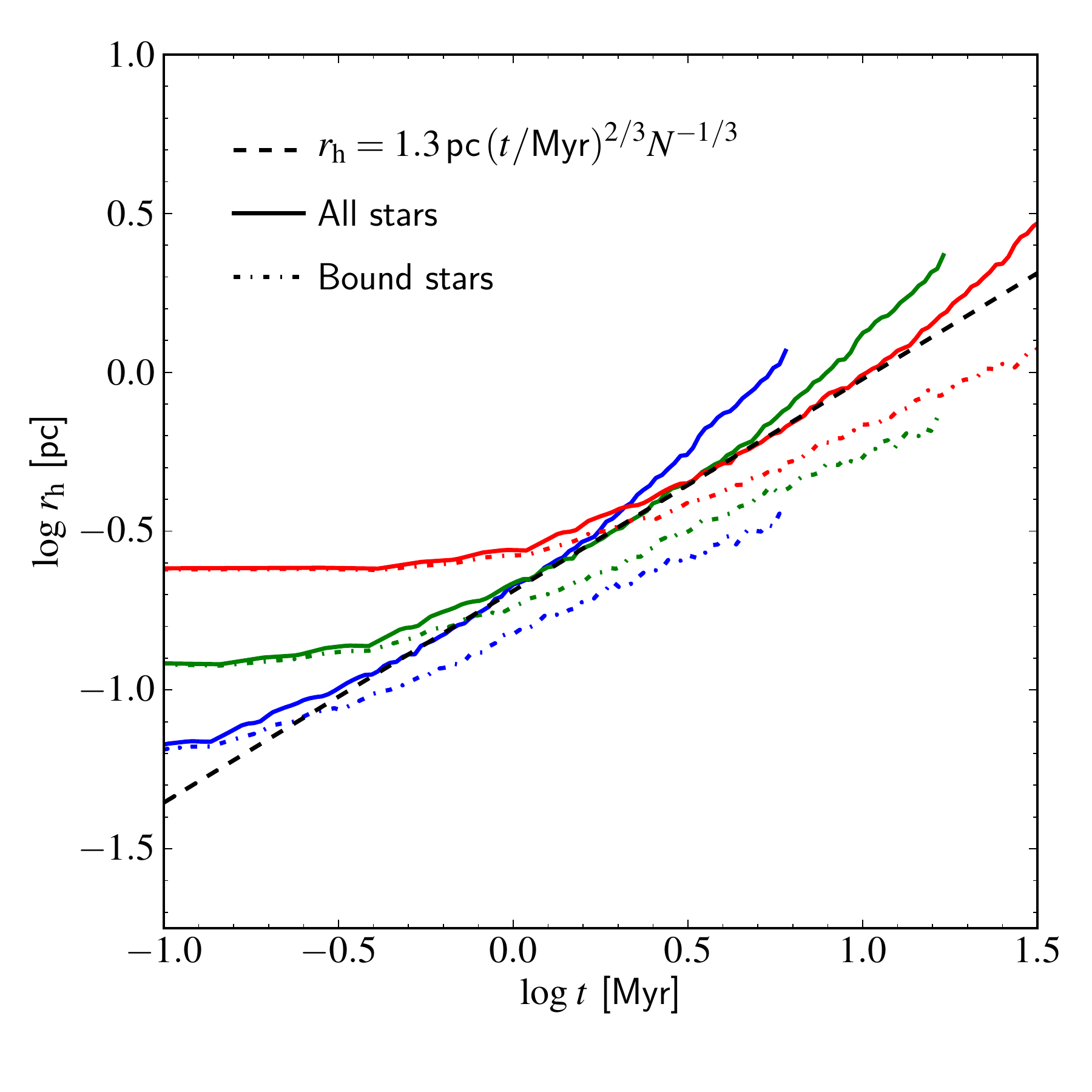} 

\caption{Evolution of the half-mass radii ($\rh$) of clusters with $N=256$ in
direct $N$-body integrations
 with {\small{NBODY}}6 \citep{2003gnbs.book.....A}. The stellar
 mass-function is a \citet{2001MNRAS.322..231K} distribution between
 $0.1\,\msun$ and $100\msun$ and the stars do not evolve internally.
 The initial  positions and velocities were generated from
 a \citet{1966AJ.....71...64K} model with $W_0=9$. In total 200
 simulations were done and the median result for $\rh$ is shown for
 different initial $\rh$. The full lines show the results when all the stars are considered and the dotted-dashed lines are for only the bound stars.
 } \label{fig:n256}
\end{figure}

Self-gravitating systems are scale free in the sense that, for a given
distribution of stars, one can re-scale their spatial separations and
not affect their qualitative evolution: in fact, scaling the spatial
separations by a factor $f$ simply involves a re-scaling of time by a
factor $f^{3/2}$.  We illustrate this in Fig.~\ref{fig:n256} where we
show the evolution of the half-mass radius for clusters of $256$ stars
and with a range of initial radii. It is evident that, although
qualitatively identical, the initially compact cluster starts its
self-similar expansion earlier. Notably, however, all three clusters
are approaching the same self-similar track by an age of a few Myr.

Another way to state this is that clusters numbering a hundred stars
end up with a mean surface density of around $\lesssim100\,\pc^{-2}$
at an age of a few Myr provided they start off with mean densities in
excess of that.  We therefore deduce that the peak of the surface
density distribution observed by Bressert et al.  is comparable to
what one would expect from dynamical evolution of stellar clusters
with ages and richness found in nearby star forming
regions \citep{2007prpl.conf..361A,2009ApJS..184...18G}.  We now turn
to considering effects which account for the finite width of the
distribution, such as the range of cluster membership and the range of
surface densities within a given cluster.

\section{The shape of the observed distribution} 
\label{sec:model}

\subsection{Surface density distribution in individual clusters}
We start by considering the distribution of surface densities in a
single cluster. To facilitate a tractable analytical approach we use
a \citet{1911MNRAS..71..460P} model to describe the surface (number)
density profile of the cluster

\begin{equation}
\sigc=\sigcn\left(1+\frac{R^2}{\rn^2}\right)^{-2}.
\label{eq:sigc}
\end{equation}
Here $R$ is the distance to the centre, $\rn$ is the scale radius and
$\sigcn$ is the central (and maximum) surface density. The latter
depends on the other variables and the total number of stars in the
cluster $N$ as $\sigcn=\ns/(\pi \rn^2)$.  The number of stars per
logarithmic surface density unit is given by
\begin{equation}
\frac{\dr n}{\dr\ln\sig} =
   \sig\frac{\dr n}{\dr R}\left|\frac{\partial R}{\partial \sigc}\right|.
\label{eq:dnds}
\end{equation}
Using $\dr n/\dr R = 2\pi R\sig$ and equation~(\ref{eq:sigc}) we find

\begin{equation}
\frac{\dr n}{\dr\ln\sig} = 
    \frac{\ns}{2}\left(\frac{\sig}{\sigcn}\right)^{1/2}, \sig\le\sigcn.
\label{eq:sigcluster}
\end{equation}
If we generalise equation~(\ref{eq:sigc}) to any cored function with
an asymptotic surface density profile with $ \Sigma \propto
R^{-\gamma}$ then the power-law slope on the right-hand side of
equation (3) becomes $1 - \gamma/2$. For the Plummer model $\gamma=4$,
hence we find a logarithmic slope of $1/2$ in
equation~(\ref{eq:sigcluster}). A shallower profile of $\gamma=3$
would give a slope of $1/3$.  Note that below a surface density of
about $\ns^{-2} \sigcn$ the expected number of stars falls to
$\lesssim1$ and we thus have $\ns^{-2}\lesssim\sig/\sigcn\le1$. For a
cluster with 100 stars there is thus a range of four orders of
magnitude in densities. It is interesting to note at this point that
the observed spread in densities in YSOs is (only) four order of
magnitude \citep{2010MNRAS.409L..54B}. This range is what one gets
from a single cluster with $N\simeq100$ stars.  In Fig.~\ref{fig:mod}
we show the $\sig$ distributions for stars in clusters with different
$N$ (dashed lines).  The sum of the distributions of clusters with
different $N$ and $\sigcn$ gives the overall $\sig$ distribution and
we will explain in  section~\ref{ssec:population} how this is
computed.

\subsection{Population of clusters}
\label{ssec:population}
To get the surface density distribution of all the stars in a
population of clusters we need to know the number of clusters with $N$
stars (i.e. the cluster initial mass function, CIMF). We adopt a
power-law scaling for the probability that a cluster has between $N$
and $N+\dr N$ stars with $N$ in the range $\nslow\le N \le \nsup$
\begin{equation}
\phic=A\,N^{-2}.
\label{eq:cimf}
\end{equation}
We can get the total $\sig$ distribution by multiplying $\phic$ by the
surface density distribution for the individual clusters
(equation~\ref{eq:dnds}) and integrating over all clusters,

\begin{equation}
\frac{\dr n}{\dr {\rm ln}\sig}=
     \int_{\nslow}^{\nsup}\phic\left.\frac{\dr n}{\dr {\rm ln}\sig}\right|_{\sigcn(N)}\dr N.
\label{eq:integral}
\end{equation}
To be able to do this, we need to relate the central surface density
$\sigcn$ to the number of stars in the cluster $N$.  Following the
arguments of Section 2 we use the size evolution of clusters of a
given age in the self-similar regime of cluster expansion. This leads
to a solution that is independent of what is assumed for the initial
mass-radius relation.  In this expansion phase the half-mass
relaxation time-scale, $\trh$, grows (roughly) linearly with time
$t$ \citep{1965AnAp...28...62H}. A constant ratio of $\trh/t$ implies
$\rn\propto N^{-1/3}t^{2/3}$ \citep{1971ApJ...164..399S}. Since the
cluster expands homologously, the central surface density is
proportional to $N/R_o^2$ and we can thus relate the central surface
density $\sigcn$ to the number of stars in the cluster and the age
as \begin{equation}
\sigcn(N,t)=\sigco \left(\frac{t}{\myr}\right)^{-4/3}N^{5/3}.
\label{eq:sign}
\end{equation}
 Here $\sigco$ is a constant of proportionality that is discussed
in section \ref{sec:proportionality}.  These relations can now be
used to perform a change of variables in
equation~(\ref{eq:integral}). The result of the integration over
$\sigcn$ is
\begin{equation}
\frac{\dr n}{\dr\ln \sig}=
          -\frac{3}{5}A\sig^{1/2}\sigcn^{-1/2}\bigg{|}_{\sigmin}^{\sigup},
\end{equation}
where $\sigup$ is the central surface density of the most massive
cluster (i.e. $\sigup=\sigco t^{-4/3} \nsup^{5/3}$,
equation~\ref{eq:sign}) and $\sigmin$ is the lower integration
boundary. There are two regimes for which different forms for
$\sigmin$ need to be considered. For $\Sigma$ smaller than the central
surface density of the smallest $N$ cluster ($\sig<\siglow=\sigco
t^{-4/3} \nslow^{5/3}$) all clusters contribute to the distribution
(see Fig.~\ref{fig:mod}). In this regime $\sigmin=\siglow$.  For
$\siglow<\sig\le\sigup$ we need to integrate only over the clusters
that contribute to the overall contribution, that is, $\sigmin=\sig$.
Using this we find for the surface density distribution of a
population of clusters

\begin{equation}
\frac{\dr n}{\dr {\rm ln} \sig}=\frac{3A}{5}\times
\begin{cases} 
\tilde{\sig}^{1/2}\left(\tilsiglow^{-1/2} -1\right)&\!\!\!, \tilde{\sig}<\tilsiglow,  \vspace{0.2cm}\\
\displaystyle1-\tilde{\sig}^{1/2}&\!\!\!, \tilsiglow<\tilde{\sig}<1, 
\end{cases}
\label{eq:dndstot}
\end{equation}
where tildes denote surface densities normalised to $\sigup$.  

The distribution has the behaviour of a single cluster
(equation~\ref{eq:sigcluster}) at low densities ($\sig\le\siglow$). In
the regime above $\siglow$ (but $<< \sigup$) the distribution is flat:
this can be readily understood since the CIMF with a logarithmic slope of $ -2$ contributes
equal numbers of stars in equal logarithmic bins of $N$ (and hence,
via equation 6, in equal logarithmic bins of $\sigc$).  The peak and
width depend on the values of $\siglow$ and $\sigup$, which both scale
with $\sigco$. So what remains to be done is to derive the constant of
proportionality   $\sigco$ that sets these values.

\subsection{Constant of proportionately $\sigco$}
\label{sec:proportionality}
To complete the model we need to know what the constant of
proportionality $\sigco$ is in the relation between surface density
and age (equation~\ref{eq:sign}). Because we assume a functional form
that goes through the origin we thereby also implicitly define the
time-scale for clusters to reach the self-similar expansion
phase. There are several factors that determine how fast clusters
reach the asymptotic track, such as the density profile, the
primordial binary star fraction, the amount of primordial mass
segregation, the initial virial ratio, etc. Our aim in this section is
to find a single value for the constant $\sigco$ that is consistent
with different numerical results.

We derive the constant of proportionality $\sigco$ from the results of
 independent sets of $N$-body simulations  of clusters
expanding under the influence of two-body relaxation. Note that these
models neglect the presence of a gaseous component. This assumption is
discussed in section~\ref{sec:conclusions}. Let us start with the
expression for the half-mass radius of an expanding cluster

\begin{equation}
\rh=\rhn \,\left(\frac{t}{\myr}\right)^{2/3} N^{-1/3}.
\end{equation}
From the $N$-body models shown in Fig.~\ref{fig:n256} we find that
$\rhn=1.3\,\pc$.  From direct $N$-body integrations of star clusters
with $8\,192\le N\le 131\,072$, a full stellar mass-spectrum and
including the effect of mass loss of the individual stars, it was
found that expansion is important already at an age of a few Myr for
clusters with initial $\trh$ less than 10
Myr \citep{2010MNRAS.408L..16G}. From their Fig.~2 an expansion of
$\rh/\rh(0)\simeq2$ was found at 1 Myr for a cluster with $N=8192$ and
an initial half-mass radius of $\rh(0)\simeq0.086\,\pc$ (the initial
$\rh$ was computed from the mass and initial half-mass density of the
cluster). This implies $\rhn\simeq3.2\,\pc$. The fact that this value
is more than a factor of 2 larger than the estimate from
Fig.~\ref{fig:n256} illustrates that this parameter is quite
uncertain. The difference is probably due to the fact that the
$N=8192$ has more massive stars compared to the $N=256$ cluster, which
speeds up the dynamical
evolution \citep{2010MNRAS.408L..16G}. \citet{2012arXiv1205.1677M}
used direct $N$-body integrations to evolve small $N$ systems (a few
hundred stars) that were selected from the outcome of  smoothed
particle hydrodynamics (SPH) simulations of a star forming molecular
cloud. From their Fig.~12 we derive $1.3\lesssim\rhn/\pc\lesssim4.5$.
The large variation in $\rhn$ might be because these clusters have
undergone varying amounts of dynamical evolution in the hydrodynamical
part of the models. Also, these clusters start off with varying
degrees of segregation of the massive stars towards the centre.
  
\begin{figure}
     \includegraphics[width=8.cm]{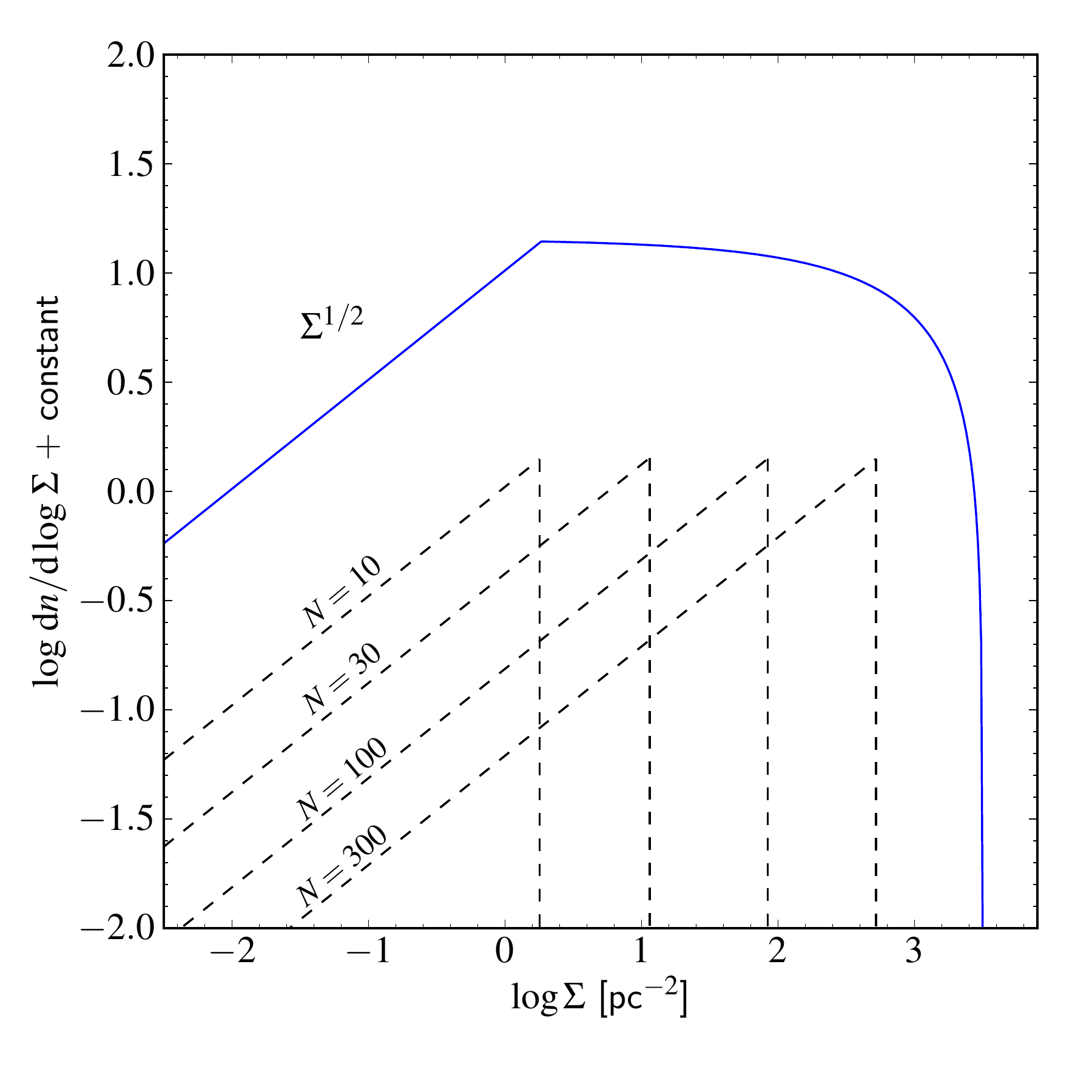} 

     \caption{Simple model for the distribution of surface densities
     around stars. All stars are in clusters that are expanding due to
     two-body relaxation. The dashed line show the surface density of
     the individual clusters with different $N$
     (equation~\ref{eq:sigc}) weighted by the CIMF. The peak density
     of each cluster is given by equation~(\ref{eq:sign}) and the sum
     (full line) follows (equations~\ref{eq:dndstot}) from integrating
     over the cluster mass function, where we used a $-2$ power-law
     distribution between 10 and 500 stars for the clusters
     (equation~\ref{eq:cimf}). An age of 2 Myr was adopted.
     } \label{fig:mod}
\end{figure}

Having compared the different results, we adopt $\rhn=2.5\,\pc$. 
We are aware that there is an uncertainty of about a factor of
two in this value.  For a Plummer model the scale radius $\rn$ is
about a factor 1.3 smaller than the half-mass radius. We can then
write $\sigco=(\pi\rn^2)^{-1}\simeq0.09\,\pc^{-2}$.  We use this 
and equation~(\ref{eq:sign}) to find $\siglow$ and $\sigup$ and to
compute the total distribution of $\sig$ using
equation~(\ref{eq:dndstot}). The result is shown in Fig.~\ref{fig:mod}
for $t = 2$ Myr. It is not exactly log-normal, but instead asymmetric
and skewed towards low densities. This is in fact something that was
found in the observations of Bressert et al. (shown in
Fig.~\ref{fig:sigma}).

The peak, or rather the flat top, of the distribution depends on the
density of the average cluster in the CIMF (Fig.~\ref{fig:mod}) and
the age of the cluster population. It shifts to lower densities as
clusters expand.  Therefore, {\it the location of the peak in this
model is insensitive to the details of the mass-size relation of
clusters at birth or the densities around the stars when they form.}
This is only true if the evolving densities of all clusters have
reached the asymptotic $\sigcn(N,t)$ relation of
equation~(\ref{eq:sign}), that is, all clusters must have been denser
in the past.  In this analytic model the peak is close to the
20\,$\pc^{-2}$ peak of the observed distribution at an age of 2
Myr. In the next section we make a slightly improved comparison by
generating the model in a Monte Carlo fashion.
 
\subsection{Monte Carlo generated model}
\begin{figure}
 \includegraphics[width=8.cm]{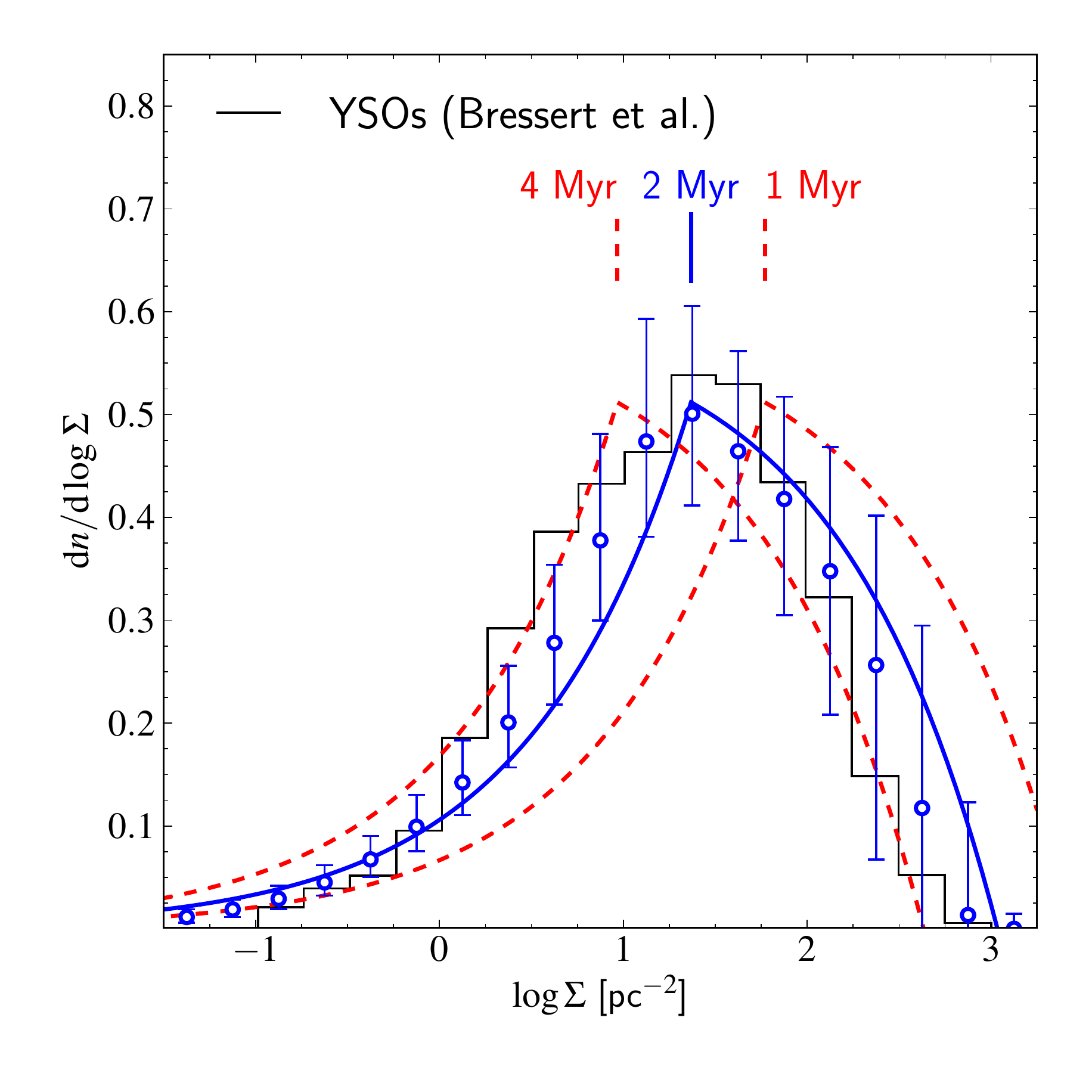} 
     \caption{Surface density distribution of YSOs (histogram) as
              found from the surveys used
              by \citet{2010MNRAS.409L..54B}.  Results of the Monte
              Carlo simulations using $50\le N\le 500$ are shown as
              open (blue) circles with error bars. The (blue) solid
              line shows the analytical model of
              section~\ref{ssec:population} for an age of 2 Myr, where
              as the (red) dashed lines show the results for ages of 1
              and 4 Myr. To conform with Bressert et al.'s figure we
              plot $\dr n/\dr \log\sig$ rather than $\log( \dr
              n/\dr \log\sig)$ as in Figure \ref{fig:mod}. All
              distributions are normalised to unit area.
              } \label{fig:sigma}
\end{figure}

Fig.~\ref{fig:sigma} shows a direct comparison between a Monte Carlo
generated sample of our model (points with error bars) to the result
for the  total of the  three Spitzer surveys of young stellar
objects (YSOs) (histogram) used in \citet{2010MNRAS.409L..54B}.  All
distributions are normalised to an area of unity.

Each cluster population contains 10 clusters with the number of stars
in the cluster randomly drawn from the mass-function of
equation~(\ref{eq:cimf}) with $\nslow=50$ and $\nsup=500$. This
results in an average of roughly $10^3$ stars in each cluster
population, which is similar to the total number of sources in the
individual  surveys considered
by \citet{2010MNRAS.409L..54B}. Note that the  Orion Nebula
Cluster (ONC) is excluded from their Orion sample and an upper limit
of a few hundred stars is,  therefore, probably appropriate for the
Solar neighbourhood minus the ONC.  For each star the surface density
is determined by finding the distance to the 7$^{\mbox{th}}$ nearest
neighbour as was done for the observations
(section~\ref{sec:intro}). An age of 2 Myr is adopted.  We generate
$1000$ cluster populations and show the median values and the
boundaries containing 67\% of the points around the median as circles
and error bars, respectively. The (blue) solid line shows the
analytical result of section~\ref{ssec:population}.

The peak values of the simulated distributions are remarkably close to
the observed peak.  We also show the analytical model results for ages
of 1 Myr and 4 Myr, to show how the peak evolves in
time. \citet{2010MNRAS.409L..54B} find a small  offset in the
density distributions of the class I and the class II objects, 
with the class II objects being slightly less dense. Because class II
objects are older this could thus be interpreted as dynamical
evolution (i.e. expansion) \footnote{Bressert et al. consider this but
conclude, in fact, that the similarity between the distribution of the
class I and class II objects supports their assumption that the
spatial distribution of YSOs is primordial.}.  There are several
physical effects we do not consider here, such as the effects of
multiple overlapping clusters and the effects of the presence of gas
on the cluster dynamics.   Although it is not our aim to develop
a model that reproduces the observations in detail  we discuss
what the effect of the presence of gas could be in the next section.

\section{Discussion and Conclusions}
\label{sec:conclusions}
We showed that the distribution of surface densities of YSOs presented
by \citet{2010MNRAS.409L..54B}, that was used to argue that only a
small fraction of the stars form in dense clusters, can be reproduced
by an extremely simple model in which all stars form in dense star
clusters. In this model the location of the peak of the distribution
is a measure of the age of the cluster population and the typical
membership number of young clusters.   Our model requires the
clusters to form with higher densities than they have at the present
day, which is a constraint on the physics of the star formation
process which can be tested with future facilities, such as the Atacama Large Millimeter/submillimeter Array (ALMA).

There are several physical effects that we have not included, 
such as the details of the spatial distribution of YSO on the sky and
the presence of gas. Although it is beyond the scope of this work to
present realistic models that include all these effects, it may be
interesting to discuss the omission of gas in our modelling in a bit
more detail. This is motivated by the observation that on the plane of the sky YSOs trace the
molecular gas they form from and the total gas mass in star forming
regions can be 10 or 20 times the total mass in
YSOs \citep[e.g.][]{2012arXiv1204.3552K}, with average gas surface
(mass) densities three or four times the surface (mass) density of
YSOs \citep{2009ApJS..184...18G}. 

A star cluster embedded in an external potential (e.g. gas) will
have a higher velocity dispersion, which slows down the two-body
relaxation that causes the expansion we discuss here. This is because
$\trh$ depends on the stellar velocity dispersion $\sigs$ and the
stellar density $\rho$ as
$\trh \propto \sigma^3/\rhos$ \citep{1971ApJ...164..399S}.  To
estimate the contribution to $\sigs$ of the gas we assume two
spherically symmetric distributions, with the same functional form for
the density profile. Using $\rs$ and $\rg$ for the half-mass radii of
the stellar and gaseous component, respectively, and $\ms$ and $\mg$
for the corresponding total masses, the velocity dispersion of the
stars can then be expressed in the stellar and gas parameters
as \citep{1969ApJ...158L.139S}

\begin{equation}
\sigs^2 \propto \frac{G\ms}{\rs} \left(1+\frac{\mg}{\ms}\frac{\rs^3}{\rg^3}\right).           
\label{eq:sig2}
\end{equation}
The first term on the right-hand side is due to self-gravity of the
stars and the second term is the contribution of the gas. If the gas
and stellar distribution have the same half-mass radius then we find
$\sigs^2 \propto (\ms+\mg)/\rs$, which is what is usually assumed for
models that consider the removal of natal gas from an embedded
cluster.  Introducing $\eta = \mg/\ms$ and $\mu = \rg/\rs$ we can
compare the relaxation time-scale of a gas embedded system to that of
a system containing only stars (i.e. what is assumed here)

\begin{equation}
\trh\mbox{(stars and gas)} = 
                 \left(1+\frac{\eta}{\mu^3}\right)^{3/2} \trh\mbox{(stars)}.
\end{equation}
From this we see that $\trh$ of a gas embedded system is longer if a
significant amount of gas is present ($\eta\gtrsim1$) which has a
comparable half-mass radius ($\mu\simeq1$). But the increase of $\trh$
depends sensitively on the size of the gaseous system in which the
stars evolve (a $\mu^{-3}$ dependence), so the effect of an additional
gas component becomes negligible if $\mu\gtrsim 3$.  Note that
$\eta/\mu^3$ is the ratio of the (volume) densities of the stars and
the gas within their half-mass radii. This is not the same as the
local (volume) density contrast.  Estimates of $\eta$ and (surface)
density contrasts are available in some dense embedded clusters but it
is not trivial to estimate the value of  $\mu$  from these
observations. If $\mu\simeq1$ the effect of dynamical expansion can be
overestimated by an order of magnitude. On the other hand, if
$\mu\gtrsim3$, our assumption of gas free cluster evolution is
justified.

From SPH simulations it was found that gas dominated star forming
regions are in fact gas-poor on the scale of the (sink)
particles \citep[i.e. $\mu>>1$,][]{2012arXiv1205.1677M}.  Arguments as
to why this might happen can be found in \citet{2012MNRAS.419..841K}.
Similar results were obtained from adaptive mesh refinement (AMR)
simulations \citep{2012MNRAS.420.3264G}. The results of these
numerical studies support our assumption of pure stellar dynamical
evolution.

We conclude that the distribution of surface densities can not be used
as evidence that not all stars form in dense clusters. Similarly, the
agreement between this model and the observations should not be
construed as an argument that {\it all} stars necessarily form in
clusters. \citet{2010MNRAS.409L..54B} argue that a (roughly)
log-normal density distribution is evidence against a scenario in
which stars form in distinct `clustered' and `distributed' star
formation modes. Their argument is that the distribution would be
bi-modal (or multi-modal) if there were distinct modes.  However, if
we interpret the model in Fig.~\ref{fig:sigma} as a `clustered' star
formation mode, we can add a `distributed` mode with a density of
several tens of YSOs per $\pc^2$ and some dispersion and the total
distribution would still be unimodal.  Our results support the
suggestion of \citet{2010MNRAS.409L..54B} that a (local) surface
density threshold is not a useful tool to separate clusters from field
stars, because in our model all stars are in clusters and there is a
range of about four orders of magnitude in surface density.

\section*{Acknowledgement}
We thank Thomas Maschberger for interesting discussions and Eli
Bressert and Nate Bastian for comments on earlier versions of the
manuscript. We also thank Simon Goodwin for a constructive referee
report.  MG acknowledges financial support from the Royal Society.

\end{document}